\documentstyle[preprint,aps]{revtex}
\tightenlines 
\begin{document}
\preprint{KNU-H016}
\def\a{\alpha}
\def\b{\beta}
\def\p{\partial}
\def\m{\mu}
\def\n{\nu}
\def\t{\tau}
\def\s{\sigma}
\def\half{\frac{1}{2}}
\def\hatt{{\hat t}}
\def\hatx{{\hat x}}
\def\hatX{{\hat X}}
\def\hatP{{\hat P}}
\def\hatth{{\hat \theta}}
\def\hatta{{\hat \tau}}
\def\hatrh{{\hat \rho}}
\def\hatva{{\hat \varphi}}
\def\p{\partial}
\def\nn{\nonumber}
\def\cb{{\cal B}}
\def\beq{\begin{eqnarray}}
\def\eeq{\end{eqnarray}}
\def\2pap{2\pi\a^\prime}

\title{String Field Theory and \\
Perturbative Dynamics of Noncommutative Field Theory}
\author{Taejin Lee
\thanks{E-mail: taejin@cc.kangwon.ac.kr}}
\address{{\it 
    Department of Physics, Kangwon National University, 
    Chuncheon 200-701, Korea}}
\date{\today}
\maketitle
\begin{abstract}
The perturbative dynamics of noncommutative field 
theory (NCFT) is discussed from a point view of string field theory. 
As in the commutative case it is inevitable to introduce a 
closed string, which may be described as a bound state of two 
noncommutative open strings. We point out that the closed 
string, interacting nontrivially with the open string, 
plays an essential role in the ultraviolet 
region. The contribution of the closed string is responsible 
for the discrepancy between the NCFT and the string 
field theory. It clarifies the controversial issues associated with the 
ultraviolet/infrared (UV/IR) behaviour of the perturbative 
dynamics of the NCFT.  
\end{abstract}


\narrowtext

\section{Introduction}

As dynamics of an open string attached to a D-brane with a $B$-field 
is known to be noncommutative \cite{con}, 
the noncommutative field theory (NCFT),
becomes a focal point in string theory recently. 
The open string theory reduces to the NCFT
in a certain limit where the massive modes decouple from the 
string dynamics. This decoupling limit has been useful to discuss
various aspects of the NCFT in the framework of string theory, 
such as the Seiberg-Witten map \cite{seib},
which relates the noncommutative Yang-Mills theory to the 
commutative one. The latest development along this avenue is
the perturbative dynamics of the NCFT \cite{min,pert,kiem}. The 
one loop amplitudes of the NCFT reveal some intriguing properties
of the NCFT, namely UV/IR mixing; its infrared behaviour strongly 
depends upon its ultraviolet behaviour. It is not easy to understand
the UV/IR mixing in the conventional renormalization scheme,
which implicitly assumes the ultraviolet-infrared decoupling.
The UV/IR mixing is considered as
an indication that the NCFT captures some novel features of 
the string theory.

As pointed out in ref.\cite{min}, the one loop effective action of the NCFT 
shows that its Wilsonian action has a undesirable feature: The 
Wilsonian action with a finite cutoff $\Lambda$ differs significantly 
from that with $\Lambda \rightarrow \infty$. In order to improve 
it, one may modify the Wilsonian action by introducing a new degree of 
freedom $\chi$. By analogy with string theory diagrams one may
interpret $\chi$ as a closed string degrees of freedom. This 
proposal has been examined as the corresponding nonplanar 
one loop amplitude of string theory is calculated. 
In ref.\cite{kiem} it has been argued that $\chi$ may be 
identified with the IR closed string degree 
of freedom if one maps the open string UV regime to the closed string 
IR regime by making use of the open/closed string channel duality. 
However, it does not fully explain 
the reason for $\chi$ to exist, since energy regimes of two channels 
do not overlap each other. The evidence that 
the closed string may play an important role in the perturbative 
dynamics of the NCFT also can be found in the NCFT at finite 
temperature \cite{closed}. 
On the contrary it has been asserted in refs. \cite{gomis1,liu} 
that no closed string modes are relevant to the UV/IR 
mixing of NCFT. Therefore, it remains an open problem to identify 
precisely the closed string degrees 
of freedom in the interacting string theory and to examine their role
in the NCFT. 

The purpose of this paper is to clarify the issue associated with
the UV/IR mixing of NCFT in the framework of the interacting string
field theory. To this end we briefly discuss construction 
of string field theory in the presence of the D-brane with a 
$B$-field. Then we show that the closed string exists
as a bound state of two open strings and its interaction
with the open string is highly nontrivial. As in the commutative
case it is inevitable to include the closed string field in
the interacting string theory: The moduli region of the nonplanar
one loop string diagram does not cover that of the corresponding 
field theory diagram \cite{kugo}. 
In order to cover the moduli region completely 
we need to introduce an open-closed string diagram. 
Thus, contribution of the closed string to the one loop nonplanar
amplitude makes discrepancy between the NCFT and the string 
theory, which may be effectively described by $\chi$. Since the
present work does not resort to the open/closed string channel 
duality, the difficulty found in the previous work is avoided.
It is clear that the string field theory serves as a complete 
framework in which the NCFT is embedded consistently.  
The relationship between the string theory and the NCFT also 
has been discussed in refs.\cite{ncstr}.

\section{Noncommutative String Field Theory}

If the D-brane carries a constant $B$-field, the open string 
variable $\hatX$ 
\beq
{\hat X}^i(\s) &=& \hat{x}^i+\theta^{ij} \hat{p}_j \left(\s -
\frac{\pi}{2}\right)+ \sqrt{2} \sum_{n=1} \left(\hat{Y}^i_n \cos n\s 
+ \frac{1}{n} \theta^{ij}\hat{K}_{j n} \sin n\s \right)
\eeq
is no longer commutative. Here
$(\hat{x}, \hat{p})$ and $(\hat{Y}_n, \hat{K}_n)$ are canonical
conjugate pairs. 
It has an important consequence that one can neither construct a 
basis for string state with eigenstates of $\hatX$, 
$\{\vert X \rangle\}$
nor express the field theoretical action in terms of the string 
functional $\Phi[X] = \langle \Phi \vert X \rangle$.
This difficulty may be overcome if we employ the basis, consisting
of the eigenstates of the momentum operators, $\{ \vert P \rangle; 
\,\,\, \hatP \vert P \rangle =P \vert P \rangle \}$. 
In contrast to $\hatX$, the momentum operators 
\beq
\hat{P}_i(\s) &=& \hat{p}_i + \sqrt{2} \sum_{n=1} \hat{K}_{in}
\cos{n\s}
\eeq
satisfy the usual commutative relation. It suggests us to construct the 
interacting string field theory on momentum space, where the 
momentum eigenstate is represented as 
\beq
\vert P \rangle &=& \exp \left[ i\int^\pi_{-\pi} d\s P(\s) \cdot
\hatX(\s) \right] \vert 0 \rangle, \\
P(\s) &=& \frac{1}{2\pi}\left(p+ \sqrt{2}\sum_{n=1} K_n \cos n\s 
\right). \nn
\eeq
With this representation we find that the inner product is given
as usual
\beq
\langle P^\prime \vert P^{\prime\prime} \rangle
= \prod_n \delta \left(P^\prime_n - P^{\prime\prime}_n\right).
\eeq

The kinetic term of the string field action can be obtained by 
evaluating the Polyakov string path integral over a strip
\beq
G(P^\prime;P^{\prime\prime}) &=& \int^\infty_0 ds \langle P^\prime
\vert \exp\left(-is \hat{L}_0 \right) \vert P^{\prime\prime} 
\rangle \\
&=& i \int D[\Phi] \Phi[P^\prime] \Phi[P^{\prime\prime}]
\exp\left[ -i\int D[P] \Phi {\cal K} \Phi \right]. \nn
\eeq
The canonical analysis given in refs.\cite{tlee} helps us to get
\beq
{\cal K} = (\2pap) \half p_i (G^{-1})^{ij} p_j + (\2pap) 
\sum_{n=1} \half \left\{ K_{in} (G^{-1})^{ij} K_{jn} - 
\frac{n^2}{(\2pap)^2} Y^i_n G_{ij} Y^j_n \right\}. \label{kine}
\eeq

Interaction term of the string field action is determined by an 
overlapping condition between three strings in momentum space.
We should note here that the usual overlapping condition in 
terms of $\hatX$ cannot be adopted, since it may not be consistent 
due to their noncommutativity.  
Let us consider the process of splitting one open string into two; the 
string 3 splits into string 1 and string 2. The quantum state of
string $I=1, 2, 3$ is represented as
\beq
\vert \Phi_I \rangle &=& \int D[P_I] 
\exp\left(i\int^{\pi\a_I}_{-\pi\a_I} d\s_I P_I(\s_I) \cdot 
\hatX_I(\s_I) \right) \vert 0 \rangle \Phi[P_I], \\
P_I(\s_I) &=& \frac{1}{2\pi\a_I} \left(p_I+ \sqrt{2} \sum_{n=1}
K_{In} \cos \frac{n\s_I}{\a_I} \right), \nn
\eeq
where $\a_1 + \a_2 = \a_3$.
The interaction term is obtained by taking the inner products 
between the strings. However, we should treat this splitting 
process with care. Before taking the inner products we need to 
rewrite the quantum state of the open string 3 as
\beq
\vert \Phi_3 \rangle &=& \int D[P_3] \exp\left(i\int_{|\s_3| \le 
\pi\a_1} d\s_3 P_3(\s_3) \cdot \hatX\right) \otimes \exp 
\left(i\int_{|\s_3| \ge \pi\a_2} d\s_3 P_3(\s_3) \cdot \hatX 
\right) \nn\\
& & \exp\left( \half \int_{|\s_3|\le \pi\a_1} d\s_3 \int_{|\s_3|
\ge \pi\a_2} d\s^\prime_3 \left[P_3(\s_3)\cdot \hatX, 
P_3(\s^\prime_3) \cdot \hatX \right] \right) \vert 0 \rangle. 
\label{split}
\eeq
The third factor in the integrand Eq.(\ref{split}) turns out 
to be the Moyal phase factor, 
$\exp\left(i \frac{\pi}{2} p_{1i} \theta^{ij} p_{2j} \right)$.
(The parameter $\theta$ in the present work differs from that
in ref.\cite{seib} by a factor of $\pi$.)
Taking the inner product we get the usual overlapping 
condition in the momentum space,
\beq
\left\{\begin{array}{l@{\quad:\quad}l}
|\s_1| \le \pi\a_1 & P_3(\s_3) = P_1(\s_1) \\
|\s_2| \le \pi\a_2 & P_3(\s_3) =P_2(\s_2) .
\end{array} \right. 
\eeq
Hence, the interaction term is determined as
\beq
V &=& \int D[P_1]D[P_2] D[P_3] \Phi[P_1] \Phi[P_2] \Phi[P_3]
\exp\left(i\frac{\pi}{2} p_{1i} \theta^{ij} p_{2j} \right) \nn\\
&& \quad \prod_{\s_1} \delta[P_1(\s_1) - P_3(\s_3)] \prod_{\s_2}
\delta[P_2(\s_2) -P_3(\s_3)].
\eeq
We may apply the same procedure also to the Witten's covariant open 
string \cite{wit}. In the low energy regime 
the string field theory effectively reduces to the NCFT.

\section{Open and Closed String Theory}

In the last section we discussed the open string field theory in 
the presence of the D-brane with a $B$-field. If we study the 
interacting open string theory, we are lead to include the closed 
string sector. The reason is that the open string theory itself is not 
unitary; the closed string would appear at higher orders 
in perturbation and the BRST invariance requires inclusion of the 
closed string. This argument also applies to the noncommutative 
case. As often discussed, the closed string degrees of freedom 
may be suppressed in the decoupling limit. At tree level this is 
certainly the case. However, at higher loop level a careful 
examination is required, since the closed string state may be
relevant in the ultraviolet regime of the loop amplitudes. 

The open string kinetic term Eq.(\ref{kine}) suggests that there exists a 
closed string with both left and right movers, which respects the 
metric $G_{ij}$. Such a closed string may be constructed as a 
bound state of the two noncommutative open strings
\beq
\vert P \rangle_{closed}  = \exp\left(i\int^{\pi}_{-\pi} d\s_1
P_1(\s_1)\cdot \hatX_1(\s_1)\right) \vert 0 \rangle \otimes
\exp\left(i\int^{\pi}_{-\pi} d\s_2
P_2(\s_2)\cdot \hatX_2(\s_2)\right) \vert 0 \rangle
\eeq
with $P_1(0) = P_2 (\pi), \quad P_2(0) = P_1(\pi)$.
Alternatively given a closed string with momentum 
$P(\s) = \frac{1}{2\pi} \sum_n P_n e^{-in\s}$, we may 
write momenta of the two open strings which constitute 
the closed string as
\begin{mathletters}
\label{mom:all}
\beq
P_1(\s) &=& \left\{\begin{array}{l@{\quad:\quad}l}
P(\s) & -\pi < \s \le 0 \\
P(-\s) & 0 < \s \le \pi
\end{array} \right. \label{mom:a} \\
P_2(\s) &=& \left\{\begin{array}{l@{\quad:\quad}l}
P(-\s) & -\pi < \s \le 0 \\
P(\s) & 0 < \s \le \pi
\end{array} \right. \label{mom:b}
\eeq
\end{mathletters}
The first equation yields
\begin{mathletters}
\label{mom1:all} 
\beq
p_1 &=& \frac{p}{2} + 
\frac{i}{2\pi}\sum_n {}^\prime P_n \left(\frac{1- 
(-1)^n}{n}\right), \label{mom1:a}\\
K_{1n} &=& \frac{1}{\sqrt{2}}\left(P_n + P_{-n}\right)
-i \frac{\sqrt{2}}{\pi} \sum_m{}^\prime P_m \left(1- (-1)^{n-m}
\right)\left(\frac{m}{n^2-m^2}\right) \label{mom1:b}
\eeq
\end{mathletters}
where in Eq.(\ref{mom1:a}) the term with $n=0$ is excluded
and in Eq.(\ref{mom1:b}) the terms with $m=\pm n$.
And the second equation yields
\begin{mathletters}
\label{mom2:all}
\beq
p_2 &=& \frac{p}{2} - \frac{i}{2\pi}\sum_n {}^\prime P_n \left(\frac{1- 
(-1)^n}{n}\right), \label{mom2:a}\\
K_{2n} &=& \frac{1}{\sqrt{2}}\left(P_n + P_{-n}\right)
+i \frac{\sqrt{2}}{\pi} \sum_m{}^\prime P_m \left(1- (-1)^{n-m}
\right)\left(\frac{m}{n^2-m^2}\right) \label{mom2:b}
\eeq
\end{mathletters}

As this closed string is formed by two noncommutative open strings
we expect that its interaction with the open string may be 
nontrivial. Since the closed string is described by two open 
strings, the interaction between the open string and the closed 
string may be obtained from the three open string interaction. It yields 
that the open-closed string interaction term acquires an 
additional phase factor, similar to the Moyal phase
\beq
\exp\left(i\frac{\pi}{2} p_{1i} \theta^{ij} p_{2j} \right)
= \exp\left[p_i \theta^{ij}\sum_n{}^\prime P_{jn} \frac{1}{4n}
\left(1- (-1)^n \right) \right]. \label{mocl}
\eeq
Here we note that higher modes of the closed string participate
in the phase factor in contrast to the open string interaction.
This phase factor would play an important role in the open/closed
string interaction in the ultraviolet regime.

\section{UV/IR Behaviours of String Field Theory}

In the low energy regime the noncommutative string field theory
(NCSFT) action reduces 
to the that of NCFT at tree level. But at higher orders in 
perturbation theory one must examine carefully the 
moduli regions which both theories cover. Here we are 
mostly concerned with the one 
loop 2-point amplitudes. A recent study \cite{kugo} on 
the one loop amplitude in 
open/closed string field theory shows that the moduli region 
covered by the nonplanar diagram is not complete. It becomes 
complete only when we take into account the open-closed-open 
string amplitude in which an open string 
becomes a closed string, then turning into the open string back. 
On the other hand the moduli region covered by the
corresponding nonplanar diagram of NCFT is complete while the
NCFT agrees well with the NCSFT in 
the (infrared) upper half part of the moduli region. 
In the (ultraviolet) low half part of the moduli region  
the open-closed-open string amplitude is relevant. 
Hence, some discrepancy between the NCFT and the NCSFT is expected.

In the commutative case the 2-point one loop tachyon amplitude 
is in the presence of $(n-1)$-brane given up to a numerical 
coefficient as \cite{green}
\begin{mathletters}
\label{amp:all}
\beq
A_{P/NP} &=&  g^2 \int d\tau dx \left[\omega^{\frac{1}{24}} f(\omega) 
\right]^{-(d-2)}(\2pap)^{-\frac{n}{2}} \left(\psi^{P/NP} (\rho,
\omega) \right)^{-2} \label{amp:a}\\
\psi^{P/NP}(\rho,\omega) &=& \frac{1\mp\rho}{\sqrt{\rho}}
\exp \left(\frac{\ln^2 \rho}{2\ln \omega}\right) \prod^\infty_{n=1}
\frac{(1\mp\omega^n\rho)(1\mp\omega^n/\rho)}{(1-\omega^n)^2} \label{amp:b}  
\eeq
\end{mathletters}
where 
$\omega = e^{-2\pi \tau}$, $ \rho = e^{-4\pi \tau x}$
and $A_{P}$ and $A_{NP}$ denote the amplitudes of planar and the 
nonplanar diagrams respectively. Note that integration
over the zero modes of momenta in the transverse directions
are not performed.
The open-closed-open string amplitude,
$A_{UU}$ takes the same expression as $A_{NP}$. 
But $A_{NP}$ and $A_{UU}$ cover different 
regions of the moduli space. 

In the noncommutative case the open string one loop amplitudes 
$A_P$ and $A_{NP}$ can be 
easily calculated if one observes that only the zero modes of momenta 
are involved in the Moyal phase and the zero momentum sectors
are factorized from the rest. It implies that the additional factor 
due to the Moyal phase is same as that of NCFT. Thus, $A_P$ does 
not receive any correction while $A_{NP}$ is modified as 
\beq
A_{NP} = g^2 \int d\tau dx \left[\omega^{\frac{1}{24}} f(\omega) 
\right]^{-(d-2)}(\2pap)^{-\frac{n}{2}} \left(\psi^{NP} (\rho,
\omega) \right)^{-2}\exp\left(- \frac{p\circ p}{\2pap \tau}\right) 
\label{np}
\eeq
where $p\circ p = -\frac{\pi^2}{4} p_i (\theta G\theta)^{ij} p_j$.
Evaluation of $A_{UU}$ in the noncommutative case would be more 
involved, since all higher modes participate in the phase 
interaction. However, in the decoupling limit the moduli of $A_{UU}$
covers mostly the ultraviolet region. Thus, it suffices to calculate 
the additional factor due to the phase interaction Eq.(\ref{mocl}) in the 
ultraviolet region. In the UV region where $\tau \rightarrow 0$ 
we may ignore the harmonic potential terms in the first quantized 
Hamiltonian of the string, so that the propagator for the 
intermediate closed string may be taken as 
\beq
{\cal K}_{closed}^{-1} = \left[(\2pap) \half \sum_{n\ge 0} 
P_{in}(G^{-1})^{ij}P_{j(-n)}\right]^{-1}.
\eeq
It greatly simplifies calculation of the additional factor due to 
the noncommutative interaction Eq.(\ref{mocl}) for $A_{UU}$
\beq
\frac{\int D^\prime[P] \exp\left( -{\cal K}_{closed}
+ p_i \theta^{ij} \sum_n{}^\prime P_{jn} \frac{1}{2n}
\left(1 -(-1)^n \right) \right)}
{\int D^\prime[P] \exp\left(-{\cal K}_{closed}\right)} 
= \exp\left(- \frac{p\circ p}{\2pap \tau}\right) 
\eeq
where the zero mode is excluded
in the integration measure $D^\prime[P]$ and we make 
use of $\sum_{n=1} 1/(2n-1)^2 = \pi^2/8$. Therefore, 
the nontrivial closed/open string phase interaction at tree level
produces the same additional factor obtained in the one loop nonplanar
open string amplitude. Taking advantage of the result obtained in the 
commutative case we conclude that the open-closed-open string 
amplitude $A_{UU}$ has the same expression as the nonplanar 
open string one loop amplitude Eq.(\ref{np}) in the 
decoupling limit. Of course, the moduli regions covered by $A_{NP}$ 
and $A_{UU}$ differ from each other. 
Now let us turn our attention to the relationship between 
the UV/IR behaviours of the NCFT and the 
NCSFT. In the decoupling limit, the 
moduli region is mostly covered by that of the nonplanar diagram. 
However, we may not completely ignore the open-closed-open string 
diagram. Let us divide the moduli region into two; region 1 where,
$\2pap\tau \ge 1/\Lambda^2$ and region 2 where, $0\le \2pap\tau <
1/\Lambda^2$. In the decoupling limit, $A_{NP}$ dominates in the 
region 1, while $A_{UU}$ in the region 2. So the contribution from 
the region 1 may be described by $A_{NP}$ as the one-loop 2-point 
function of the NCFT with a cutoff $1/\Lambda^2$ in the decoupling
limit. The contribution from the region 2
may well be approximated by $A_{UU}$, the integral over the moduli
region 2, where $s=1/\tau$, $s/(\2pap) > \Lambda^2$, which 
may be written in the decoupling limit as
\beq
A_{UU} \simeq \int^\infty_0 \frac{ds}{s} s^{\frac{n-24}{2}}
\left(\exp\left(-\frac{p\circ p \,s}{\2pap}\right) -
\exp\left(-\frac{\left(p\circ p + 1/\Lambda^2 
\right) s}{\2pap}\right) \right)
\eeq
It can be reproduced as the one-loop 
amplitude by the following field theoretical effective action 
\cite{min} which has been introduced to improve the
Wilsonian action for the NCFT
\beq
\int L_{eff} = \int d^{26} x \chi \phi + \int d^{26} x \left(
\p \chi \circ \p \chi + \Lambda^2 (\p \circ \p \chi)^2 \right),
\eeq
where $\phi$ is the noncommutative scalar field, 
describing the open string degrees of freedom. 
It has been suggested that the
extra degrees of freedom, $\chi$ field denotes the closed string 
degree of freedom at the low energy regime and the metric 
respected by the closed string is $\theta G \theta$. However,
the present analysis reveals that the $\chi$ field itself
is not simply related to
the closed string field and the closed string does not respect 
the metric $\theta G \theta$, but respects the metric $G$ like the
open string.

\section{Conclusions}

In the present paper we construct an interacting NCSFT and 
compare its perturbative dynamics with that of the NCFT. 
It is inevitable to introduce the closed string field 
in the interacting open string theory as in the case of the commutative 
case and its contributions to the string amplitudes do not vanish even in 
the decoupling limit. 
The one-loop 2-point tachyon amplitudes have been calculated in the
literature in the first quantized string theory. The works of 
refs.\cite{min,kiem,closed} suggest that the closed string may 
play an important role in the perturbative dynamics of the NCFT.
On the other hand the works of refs.\cite{gomis1,liu} asserts that the
closed string modes are irrelevant to the UV/IR mixing of 
the NCFT and the $\chi$ field cannot be realized in the string 
theory. However, we note that the one-loop 2-point tachyon 
amplitudes have been discussed only in the first quantized string
theory in the literature, although the second quantized string
theory is more suitable to examine the role of the closed string.
In the interacting open string theory a closed string can be 
formed as a bound state of two open strings. We find that the 
closed string, being formed as a bound state of two noncommutative
open strings, interacts with the open strings nontrivially and
their contribution to the 2-point tachyon amplitudes is
described by the $\chi$ field in the decoupling limit.
This closed string cannot be described by the first quantized 
string theory which adopts the Green function for the open string 
only. The amplitude of the nonplanar diagram reproduces 
the corresponding amplitude of the NCFT in the decoupling limit
as discussed in the literature.
But it does not imply that the closed string cannot
contribute to the 2-point tachyon amplitude. According to a
detailed study of the 2-point tachyon amplitudes\cite{kugo}
the amplitude of the nonplanar diagram $A_{NP}$ and 
the open-closed-open string amplitude $A_{UU}$ cover different
regions of the moduli space. Thus, $A_{UU}$ should be taken 
into account in addition to $A_{NP}$ when we evaluate the 2-point 
tachyon amplitude. In the decoupling limit, $A_{NP}$
is approximated by the one-loop amplitude of the NCFT
with a cutoff and $A_{UU}$ can be described as the contribution 
of the $\chi$ field.

The closed string is found to contribute to the perturbative 
dynamics of the NCFT in a more intriguing way than we expect. 
It respects the metric $G$ like the open string, but its 
interaction with the open string is nontrivial. 
The string field theory serves as a proper framework 
where the perturbative dynamics 
of the NCFT can be discussed consistently. It is expected that the 
NCSFT also greatly improve our understanding of
the renormalizability of the NCFT and 
have important applications in other interesting subjects 
of the NCFT. 


\section*{Acknowledgement}
This work was supported by KOSEF (995-0200-005-2). 
Part of the work was done during the author's visit to APCTP (Korea), 
NBI (Denmark) and KIAS (Korea). I would like to thank J. Ambjorn 
G. Semenoff, and Y. Makeenko for inviting me to NBI.

\end{document}